\documentstyle[11pt,aaspp4,psfig,flushrt]{article}

\def\la{\mathrel{\hbox{\rlap{\hbox{\lower4pt\hbox{$\sim$}}}\hbox{$<$}}}}
\def\ga{\mathrel{\hbox{\rlap{\hbox{\lower4pt\hbox{$\sim$}}}\hbox{$>$}}}}

\begin{document}

\lefthead{COORAY, QUASHNOCK, \& MILLER}
\righthead{LOWER LIMIT ON $\Omega_m-\Omega_{\Lambda}$ FROM THE HDF}

\title{A Lower Limit on ${\bf \Omega_m-\Omega_{\Lambda}}$
Using Gravitational Lensing in the Hubble Deep Field}
\author{Asantha R. Cooray, Jean M. Quashnock, M. Coleman Miller}
\affil{Department of Astronomy and Astrophysics,
University of Chicago, Chicago IL 60637\\
E-mail: asante@hyde.uchicago.edu, jmq@oddjob.uchicago.edu,
miller@bayes.uchicago.edu}

\received{1998 June 5}
\revised{1998 August 14}
\accepted{1998 September 9}

\slugcomment{To appear in {\em The Astrophysical Journal}}

\begin{abstract} We calculate the expected number of multiply-imaged
galaxies in the Hubble Deep Field (HDF), using photometric redshift
information for galaxies with $m_I < 27$ that were detected
in all four HDF passbands.
A comparison of these expectations with
the observed number of strongly lensed galaxies places a lower limit on
the current value of $\Omega_m-\Omega_{\Lambda}$, where $\Omega_m$ is
the cosmological mass density of the universe and $\Omega_\Lambda$ is
the normalized cosmological constant. Based on current estimates of the
HDF luminosity function and associated uncertainties in individual
parameters, our 95\% confidence lower limit on
$\Omega_m-\Omega_{\Lambda}$ is between -0.44, if there are no strongly
lensed galaxies in the HDF, and -0.73, if there are two strongly
lensed galaxies in the HDF. If the only lensed galaxy in the HDF is the 
one presently viable candidate, then, in a flat universe
($\Omega_m+\Omega_\Lambda=1$), $\Omega_{\Lambda} < 0.79$ (95\% C.L.).
These lower limits are compatible with estimates
based on high-redshift supernovae and with previous limits
based on gravitational lensing.

\end{abstract}



\section{INTRODUCTION}

The Hubble Deep Field (HDF) is the deepest optical survey
that has been made to date, enabling a detailed study of the
galaxy redshift distribution (e.g., Gwyn \& Hartwick 1996) and the
global star formation history (e.g., Madau et al. 1996, 1998). The HDF covers
an area of 4.3 arcmin$^{2}$, with a pixel scale of $0\farcs04$,
and has 5 $\sigma$ point source detection thresholds of 26.7, 28.8, 30.1,
and 30.3 mag measured in the Space Telescope magnitude system
in F300W, F450W, F606W, and F814W filters, respectively
(e.g., Clements \& Couch 1996).
In this paper, we use AB magnitudes;
in this system, the magnitudes of the detection thresholds are,
respectively, 28.0, 29.2, 29.9, and 29.5 mag.
Complete observational and data reduction
details of the HDF are given by Williams et al. (1996).

Spectroscopic redshifts exist for nearly 180 galaxies  in the
HDF.  These redshifts are now complemented by two photometric redshift
catalogs,  one based on spectral template fitting (Sawicki, Lin, \& Yee
1997; hereafter SLY), and the other on empirical color-redshift relations
(Wang, Bahcall, \& Turner 1998; hereafter WBT).
Here we use the magnitudes for HDF sources derived by SLY
and the redshifts derived by SLY and WBT.
Galaxies in the HDF have redshifts which are estimated to range from 0.1
to 5, with a large portion having redshifts between 2 and 4.
Such galaxies have a significant probability of being strongly lensed.
The Hubble Space Telescope (HST) has proven to be invaluable
to gravitational lens discovery programs because of the
high resolution images it produces (e.g., Ratnatunga et al. 1995). The
combination  of high resolution and deep exposures in multiple colors
provides a rich ground for gravitational lens searches,
and it was expected that HDF would contain between 3 to 10 lensed galaxies,
based on the number of lensed quasars and radio sources in other surveys
(Hogg et al. 1996).

Instead, a careful analysis of the HDF (e.g., Zepf et al. 1997) has
revealed a surprising dearth of candidates for lensed sources. In fact,
the best estimate is either 0 or 1 lensed sources in
the entire field,  although very
faint images with small angular  separations may have escaped current
analyses.  This lack of lensing has led to suggestions (e.g., Zepf et
al. 1997) that the HDF data may be incompatible with the high probability
of lensing expected in a universe with a large cosmological constant.

Here, we calculate the expected number of detectable,
multiply-imaged galaxies in the HDF
for different cosmological parameters, and we constrain
these parameters by comparing the expectations with the observations.
In \S~2 we discuss our calculation and its inputs,  including
the inferred redshifts of the galaxies and the luminosity  distribution
of potential lenses.  In \S~3 we present our resulting constraints on
cosmological parameters, and in particular on
$\Omega_m-\Omega_\Lambda$.  In \S~4 we discuss the potential effect of
systematic errors on the expected and observed number of lensed galaxies.
Finally in \S~5 we summarize and discuss future prospects for
tighter constraints. We follow the conventions that the Hubble
constant, $H_0$, is 100\,$h$\ km~s$^{-1}$~Mpc$^{-1}$, the present mean
density in the universe in units of the closure density is $\Omega_m$,
and the present normalized cosmological constant is $\Omega_\Lambda$.
In a flat universe, $\Omega_m+\Omega_\Lambda=1$.

\section{EXPECTED NUMBER OF LENSED GALAXIES IN THE HDF}

In this section we describe our calculation of the expected number of
lensed galaxies in the HDF.
In \S~2.1 we discuss the formalism for the calculation, and in
\S~2.2 we consider input quantities and their errors, such as the
photometric redshifts and the luminosity distribution of potential
lenses.

\subsection{Calculating the Number of Lensed Galaxies}

In order to calculate the number of lensed galaxies in the HDF, we model the
lensing galaxies as singular isothermal spheres (SIS) and use the
analytical  filled-beam approximation (e.g., Fukugita et al.
1992). At redshifts $z \lesssim 4$, the analytical filled-beam
calculations  in Fukugita et al. (1992) agree to better than 2\% with
numerical calculations (e.g.,  Fig.\ 1 in Holz, Miller, \&
Quashnock 1999).

To calculate the probability of observing strong lensing,
two additional effects must be included.
First, any magnitude-limited sample such as the HDF is
subject to so-called ``magnification bias" (e.g., Kochanek 1991), in
which the number of lensed sources in the sample is
larger than it would be in an unbiased sample,
because lensing brightens into the sample sources
that would otherwise not be detected.
This is a particularly pronounced effect in quasar lensing
surveys (e.g., Maoz \& Rix 1993), because the faint end of the
quasar luminosity function rises steeply.  Second,
because identification of a lensed source requires the detection of at
least two of its multiple images, then if the second brightest of the images
is too faint to detect, such sources will not be identified as strongly
lensed. For lens-search surveys such as the HST Snapshot
Survey of bright quasars (Bahcall et al. 1992),  this effect is small,
because the depth of the pointed search is much greater than the depth
of the initial search, and hence almost all secondary images will be
detected.  In the case of the HDF, it is not possible to do a deeper
follow-up, and hence for a significant fraction of sources, especially
those near the limiting magnitude of the survey, it will be difficult to
detect other lensed companions.

If the probability for a source at redshift $z$ to be strongly lensed
is $p(z,\Omega_m,\Omega_\Lambda)$,
and the number  of unlensed sources in the HDF between rest-frame
luminosity $L$ and  $L+dL$ and between redshifts $z$ and $z+dz$ is
$\Phi(L,z)dL\,dz$, then the number of lensed sources, $d\bar N$, in that
luminosity
and redshift interval expected in the HDF is (see also Maoz et al.\
1992)

\begin{equation}
{d\bar N(L,z)\over dz}=p(z,\Omega_m,\Omega_\Lambda)
\int\left[\Phi\left({L\over A},z\right)\,
{dL\over A}\right]f(A,L,z)q(A)\,dA
\end{equation}
where the integral is over all allowed values of $A$, the amplification
of the brightest lensed image compared to the unlensed brightness,
$q(A)$ is the probability distribution of amplifications, and
$f(A,L,z)$ is the probability of observing the second-brightest image
given $A$, $L$, and $z$.  Our assumption that the lenses are singular
isothermal spheres implies that the minimum amplification is $A_{\rm
min}=2$ and the probability distribution is
$q(A)=2/(A-1)^3$.  For simplicity, we will assume that
$f(A,L,z)$ is a step function ($\Theta$), so that a
dimmer image with apparent magnitude brighter than $m_{\rm lim}$
is detected, whereas one dimmer than $m_{\rm lim}$ is not detected.
This assumption means that

\begin{equation}
f(A,L,z)=\Theta\left[m_{\rm lim}-m_i-2.5\log_{10}
\left(A\over{A-2}\right)\right]\; ,
\end{equation}
where $m_i$ is the apparent magnitude of the brighter image and
$A/(A-2)$ is the ratio of the brightness of the primary image
to the brightness of the secondary image in the SIS approximation.

We assume that the brightness distribution of galaxies
at any given redshift is described by a Schechter function, in
which the comoving density of galaxies at redshift $z$ and with
luminosity between $L$ and $L+dL$ is

\begin{equation}
\phi(L,z)\, dL=\phi^*(z)\left[L\over{L^*(z)}\right]^{\alpha(z)}
e^{-L/L^*(z)}\, dL\; ,
\end{equation}
where, as before, both $L$ and $L^*$ are measured in the rest
frame of the galaxy.
Thus, $\Phi(L,z) = \phi(L,z)\, dV/dz$,
where $V$ is the comoving volume in the solid angle
of the HDF. We can then write the
expected number $\bar N$ of lensed galaxies in our selected subsample
of the HDF as

\begin{eqnarray}
\bar N &=\sum_i p(z_i,\Omega_m,\Omega_\Lambda)
\int_2^\infty A^{-1-\alpha(z_i)}e^{L_i/L^*(z_i)}
e^{-L_i/AL^*(z_i)}\nonumber \\
&\times \Theta\left[m_{\rm lim}-m_i-2.5\log_{10}\left(A\over{A-2}\right)\right]
{2\over{(A-1)^3}}dA\nonumber \\
&\equiv \sum_i \tau(z_i) \; .
\end{eqnarray}
Here the sum is over each of the galaxies in our sample,
where we have chosen only those galaxies in the HDF with
I magnitudes brighter than 27.  The index $i$ represents
each galaxy; hence, $z_i$, $L_i$, and $m_i$ are, respectively,
the redshift, rest-frame luminosity, and apparent I magnitude
of the $i$th galaxy.

The quantities $z_i$ and $m_i$ can be measured with relatively
little error, but the rest-frame luminosities $L_i$ and $L^*(z_i)$
are more difficult to infer because their value depends
on uncertain K-corrections.  Therefore, we estimate the
total average bias by summing the expectation values of $\tau(z_i)$,
which are computed by weighting the integral in equation (4) by a normalized
distribution of luminosities $L_i$ drawn from the Schechter function
appropriate for the redshift $z_i$ of galaxy $i$,
instead of using inferred rest-frame luminosities that
are, in any case, very uncertain.
We used the tabulated values for Schechter luminosity
function in SLY, and calculated the magnification
bias for individual redshift intervals for which the Schechter function
parameters are available in Table 1 of SLY.  In principle, the uncertainties
in the Schechter function parameters at a given redshift can affect
the calculation of the bias, but in practice only the
uncertainty in the power-law slope $\alpha$ has a significant effect;
in \S~3.2 we calculate the result of varying $\alpha$ by the quoted
errors for source galaxies with redshifts between 2 and 3, which
dominate the expected incidence of lensing.

\subsection{Uncertainties in Inputs}

{\em Redshifts of galaxies}---We have used
two available photometric redshift catalogs for galaxies
in the HDF: SLY used spectral fitting techniques to
calculate the redshift for all galaxies that appeared in the four HDF
passbands with a I-band limiting magnitude of 27. There are 848
such galaxies, 181 of which have spectroscopic redshifts.
Recently, WBT computed photometric redshifts for the
same sample as SLY, based on empirical relations
which were calibrated against spectroscopic redshifts.  In Figure~1,
we show the redshift distribution of the HDF galaxies according to the
SLY (left panel) and WBT (right panel) catalogs,
as a function of the I-band magnitude. These distributions
are similar to that found by Gwyn \& Hartwick (1996), and
contain  two peaks, with one at  $z \sim 0.6$ and the other at $z\sim$
2.3. Compared to the spectroscopic redshifts, the photometric redshifts
of SLY have a larger scatter than the redshifts
of WBT.  However, as is clear from Fig.\ 1, there is a
pronounced (and most probably spurious)
lack of photometric redshifts between 1.5 and 2.2 in the WBT
catalog, in the same redshift range
where no spectroscopic redshifts are currently available for the HDF.
This gap is much less dramatic in the SLY catalog.
It is therefore uncertain which catalog is more
reliable overall.  We therefore estimate cosmological constraints based
on both catalogs, but we find (see \S~3) that the derived constraints
are almost the same for either catalog.

{\em Properties of lensing galaxies}---The overall probability of
strong lensing depends on the number density and
typical mass of lensing galaxies.  For singular isothermal spheres,
with comoving number density $n_0$ and velocity dispersion $\sigma$,
this factor is given by the dimensionless parameter
\begin{equation} F\equiv
16\pi^3n_0R_0^3\left(\sigma\over{c}\right)^4\; .
\end{equation}
where $R_0\equiv c/H_0$.  The
parameter $F$ is independent of the Hubble constant, because the
observationally inferred  number density is itself proportional to $h^3$.
Note that the probability of lensing, $p(z,\Omega_m,\Omega_\Lambda)$,
is directly proportional to $F$.

At any given redshift, we can estimate $F$ directly from the galaxy luminosity
function at that redshift, if we know the dependence of the velocity
dispersion on the luminosity.  A commonly assumed functional form is
$L\propto \sigma^\gamma$, so that $\sigma^4\propto L^{4/\gamma}$.
Kochanek (1996) estimates $\gamma=4.0\pm 0.5$ and adopts a velocity
dispersion for an $L^*$ galaxy in the local universe of
$\sigma = 220 \pm 20$ km s$^{-1}$.  If the luminosity
function at redshift $z$ is given by equation (3), then with the
normalization for the velocity dispersion given by Kochanek (1996), we
find, by integrating equation (5) over the luminosity function at
redshift $z$, that

\begin{equation}
F=3.87\left[L^*(z)\over{L^*(0)}\right]^{\frac{4}{\gamma}}
\phi^*(z)\Gamma\left(\alpha+\frac{4}{\gamma}+1\right)\; ,
\end{equation}
where $\Gamma$ is the normal gamma function.  We assume that the
luminosity at which $\sigma=220$ km s$^{-1}$, corresponds to a B
magnitude of $M^*_B=-20.7$.  Henceforth, we will also assume $\gamma=4$.

To estimate $F$, we concentrate on the redshifts $z$ between 0.5 and 1.0,
because foreground lensing galaxies for most of the presently confirmed
lensed sources are in this redshift range (e.g., Kochanek 1996).
SLY find that in this redshift range, the HDF
luminosity function for galaxies is represented by a Schechter
function with the following parameters:
\begin{equation}
M^*_B=-19.9\pm 0.3, \alpha=-1.3\pm 0.1,
\phi^*=0.042\pm 0.013\ h^3{\rm Mpc}^{-3}
\end{equation}
The best estimate for $F$ is therefore 0.05, if 30\% of the galaxies
in the HDF are ellipticals.  If the errors in the
parameters were independent of each other, the  uncertainty in $F$
would be at least a factor of 2.  However, the parameters in the Schechter
luminosity function are  correlated.  For example, in the Century
Survey (Geller et al.\ 1997) the joint error ellipses for $\alpha$ and
$M^*$ show that, in the R band, $\alpha\approx -1.2+(M^*_R+20.7)$.  It
is therefore plausible that a similar relation holds for the HDF
luminosity function parameters.
>From Table~1 of SLY, we find that, for $z$ between 0.5 and
1.0, $\alpha\approx -1.3 + 0.3(M^*_B+19.9)$.   In addition, the
absolute normalization is well-determined at the faint end of the
luminosity function, where there are many galaxies.  If we fix the
number density at $0.01L^*$, where galaxies are numerous yet still
bright enough to be detected reliably, then a variation in $\alpha$
determines the values of $M^*_B$ and $\phi^*$.  The 1$\sigma$ range in
$\alpha$ then gives values of $F$ in a tight range, between 0.100 and
0.108.  However, the correlation between $\alpha$ and $M^*_B$
is inexact; furthermore, other uncertainties,
such as in the velocity dispersion $\sigma$, must also be considered.
To be conservative and account for systematic uncertainties
in the form of the $\sigma$-$L$ relation
and in the joint errors of the Schechter parameters,
we allow for an overall uncertainty of 30\%, and take $F=0.050\pm 0.015$.

\section{CONSTRAINTS ON COSMOLOGICAL PARAMETERS}

\subsection{Observed Number of Lensed Galaxies in the HDF}

The first potential detection of a gravitationally lensed source in the
HDF is described in Hogg et al. (1996). However, the candidacy of this
object was later questioned by Zepf et al. (1997), who made Keck
spectroscopic observations of 3 lens-like objects in the HDF.  These
three sources were selected based on the morphological distribution of
nearly 750 galaxies, down to a limiting magnitude of 27 in the I-band.
Discrepancies in the inferred redshifts of the multiple images
suggested that  two of the three candidates are not lensed sources,
leaving one possible gravitational lens in the HDF down to I of 27
(L3.2 in Zepf et al. 1997). It is possible that even this source is not
strongly lensed, but currently the data are inconclusive.  Zepf et al.
(1997) argue that it is unlikely that there is a large population of
lensed sources in the HDF that has been missed.  This is especially
true for sources brighter than $m_I=27$, 1.5 mag above the detection
threshold, $m_{\rm lim}$, of 28.5 in the I-band.   However, in order to
take into account the possibility that multiply-imaged sources with
faint secondary images (which still have $m_I<28.5$) or small angular
separations have been missed, we calculate cosmological limits based on
the assumed detection of 0, 1, or 2 lensed sources with secondary images
brighter than $m_{\rm lim}=28.5$, $m_{\rm lim}=28.0$, and $m_{\rm
lim}=27.5$ in  the HDF.  As we discuss in  more detail in the next
section, the limits that follow from the detection  of at least one
lensed source require knowledge of the redshift of the background
lensed source.  The single candidate of Zepf et al.\ (1997) has an
uncertain redshift, but the redshift is most probably in the range $1.0
\lesssim z \lesssim 2.5$, with a best guess of 1.02.

\subsection{Comparing the Expected with the Observed Number of Lensed Galaxies}

We have calculated the expected number, $\bar N$,
of detectable, multiply-imaged galaxies in
the HDF, using equation (4) for various combinations of $\Omega_m$,
$\Omega_\Lambda$,  $m_{\rm lim}$, and $\alpha$,
the luminosity function power-law
slope for galaxies with redshifts between 2 and 3
(the redshift range expected to produce most of the lensed galaxies in the
HDF).
We perform each calculation using both the SLY
and WBT catalogs. Figure~2 shows the expected number of
gravitational lenses in the HDF as a function of $\Omega_m$ and
$\Omega_{\Lambda}$, assuming $m_{\rm lim}=28.5$ and $\alpha=-2.1$,
and using the SLY photometric redshifts.
A universe dominated with $\Omega_\Lambda$ has a higher number of
multiply-imaged
sources than in a universe dominated with a large $\Omega_m$.
As shown in Fig.\ 2,
$\bar N$ is essentially a function of the combined quantity
$\Omega_m-\Omega_{\Lambda}$. This degeneracy in the lensing probability
(Carroll, Press, \& Turner 1992; Kochanek 1993;
Holz, Miller, \& Quashnock 1999)
allows us (see below) to constrain $\Omega_m-\Omega_{\Lambda}$
rather than $\Omega_m$ or $\Omega_{\Lambda}$ individually.
In Table 1,
we list the expected number of strongly lensed galaxies in the HDF along
the  $\Omega_m+\Omega_\Lambda=1$ line  as a function of
$\Omega_m-\Omega_\Lambda$. These expected numbers, which are listed for the
two catalogs separately, have been calculated assuming $m_{\rm
lim}=28.5$ and $\alpha=-2.1$.

\subsection{Likelihood Constraints on ${\bf \Omega_m - \Omega_\Lambda}$}

We constrain the quantity $\Omega_m - \Omega_\Lambda$
by comparing the observed and expected number of
lensed galaxies in the HDF.
We adopt a Bayesian approach, and take a prior for
$\Omega_m - \Omega_\Lambda$ that is uniform between -1 and +1.
We do this because we do not yet have a precise determination of
this quantity (although recent high redshift supernovae
measurements favor a value near -0.5; e.g., Riess et al. 1998),
and we do not wish to consider cosmologies in
which either $\Omega_m$ or $\Omega_\Lambda$ lie outside the interval [0,1].
We do not constrain $\Omega_m$ or $\Omega_\Lambda$ separately;
thus, no prior is required for these quantities.
Since the prior for $\Omega_m - \Omega_\Lambda$ is uniform,
the posterior probability density is simply proportional to the
likelihood (see below).

>From equation~(4), $\bar N = \sum_i \tau(z_i)$,
where $\tau(z_i)$ is the effective probability that a galaxy at
redshift $z_i$ is lensed.
Here, $\tau(z_i)$ and $\bar N$ depend on
$\Omega_m - \Omega_\Lambda$, and are directly proportional
to the dimensionless parameter $F$ defined in equation~(5).
Furthermore, we take into account the uncertainty in $F$
by defining ${\cal F} \equiv F/0.05$ and taking ${\cal F}$
to have a mean of unity
and standard deviation $\sigma_{\cal F}=0.3$.
Thus, we allow for a 30\% uncertainty in $F$ (see \S\ 2).
The factor ${\cal F}$ is then an overall correction to the
lensing probability, due to a systematic uncertainty in $F$.

The likelihood ${\cal L}$ --- a function of $\Omega_m - \Omega_\Lambda$ ---
is the probability of the data, given $\Omega_m - \Omega_\Lambda$.
If there are no lensed galaxies in the HDF,
and if ${\cal F}$ were known exactly, this probability would be
\begin{eqnarray}
{\cal L}(0) & = & \prod_i e^{-{\cal F} \tau(z_i)}
	= e^{-{\cal F} \bar N}\nonumber \\
       & = & e^{-\bar N} \times e^{-({\cal F} - 1) \bar N} \;  .
\end{eqnarray}
We account for the uncertainty in $F$ by marginalizing the  above expression
over ${\cal F} = 1.0 \pm \sigma_{\cal F}$, expanding the exponential,
and taking expectation values. To second order in $\sigma_{\cal F}$,
we find
\begin{equation}
\langle {\cal L}(0)\rangle = e^{- \bar N} \times
                \left(1 + \sigma_{\cal F}^2 \frac{\bar N^2}{2}\right) \; .
\end{equation}
We only include contributions from the variance in ${\cal F}$
to the expectation value, since we do not know the exact
distribution (and higher moments) of ${\cal F}$.

If instead there is one lensed galaxy (at redshift $z_j$) in the HDF,
and if ${\cal F}$ were known exactly, the likelihood would be
\begin{eqnarray}
{\cal L}(1) & = & {\cal F}\tau(z_j)\times \prod_i e^{-{\cal F} \tau(z_i)}
        = \tau(z_j)\times {\cal F} e^{-{\cal F} \bar N} \nonumber \\
    & = & \tau(z_j)\times\left(-{\frac{\partial {\cal L}(0)}{\partial \bar
N}}\right)\; .
\end{eqnarray}
By again marginalizing over ${\cal F}$, and substituting equation (9) into
equation (10), we obtain
\begin{equation}
\langle {\cal L}(1)\rangle = \tau(z_j)\times e^{- \bar N} \times
        \left(1 + \sigma_{\cal F}^2 \left[ \frac{\bar N^2}{2}-\bar
N\right]\right) \; .
\end{equation}
In general, if there are $n$ lensed galaxies (at redshifts $z_j$)  in the HDF,
then we find
\begin{equation}
\langle {\cal L}(n)\rangle =\prod_{j=0}^n \tau(z_j) \times e^{- \bar N} \times
\left(1 + \sigma_{\cal F}^2 \left[ \frac{\bar N^2}{2}-n\bar N
+\frac{n(n-1)}{2}\right]\right) \; .
\end{equation}

In order to constrain $\Omega_m - \Omega_\Lambda$,
we calculate the likelihood for cases in which the
number $n$ of lensed galaxies present in the HDF is
0, 1, or 2, and for $m_{\rm lim}$=28.5, 28.0, and 27.5.
In order to examine the effect of the luminosity function slope $\alpha$ ---
which has an important effect on magnification bias (eq.~[4]) ---
we have also calculated the
likelihood by varying $\alpha$ by $\pm$ 0.1 (the quoted error in SLY)
from the best estimate of -2.1
(for galaxies with redshifts between 2 and 3, which dominate the
expected incidence of multiple imaging).

In Table 2, we present the 95\% confidence lower limits on
$\Omega_m-\Omega_\Lambda$ for various cases.
We define a canonical case in which the lens search has been carried out
to $m_{\rm lim}=28.5$, and has found one lensed galaxy ($n=1$)
in the HDF with a redshift $z_s=2.5$. We take $\alpha=-2.1$ in the canonical
case.
In order to test the effect of different number of
lensed galaxies observed in the HDF, we vary $n$ from
this canonical scenario and assume that all lensed
galaxies are at a redshift of 2.5
(this gives the weakest lower limit).
The cumulative probabilities for the observed number of lenses,
as a  function of $\Omega_m-\Omega_{\Lambda}$,
are shown in Figure~3,  where the
plotted curves represent no lensed galaxies in HDF (solid), one lensed
galaxy (short-dashed) and two lensed galaxies (long-dashed).
If there are no lensed galaxies in the HDF, then
at the 95\% confidence level $\Omega_m-\Omega_{\Lambda} > -0.44$, so
that in a flat universe  $\Omega_{\Lambda} < 0.72$. If there is one
lensed galaxy in the HDF, our constraints depend only slightly on the galaxy
redshift.  If the galaxy redshift is 1, then $\Omega_m-\Omega_{\Lambda}
> -0.52 $, implying $\Omega_{\Lambda} < 0.76$ in a flat universe. If
instead the galaxy redshift is 2.5, then  $\Omega_m-\Omega_{\Lambda} > -0.58
$, and hence $\Omega_{\Lambda} < 0.79$ in a flat universe.

As tabulated in Table 2, the change in I-band  lens-search magnitude
limit from 28.5 to 27.5  has a surprisingly small effect on the limits on
$\Omega_m-\Omega_\Lambda$;  the effect of the nondetection of secondary
images beyond the limiting magnitude is compensated to some extent by
the effect of magnification bias. Except in the case where two strongly
lensed galaxies are present in the HDF, $\Omega_m-\Omega_\Lambda >
-0.70$ at the 95\% confidence level. This implies that in a flat
universe $\Omega_\Lambda < 0.85$, which is  consistent with the  recent
cosmological parameter constraints based on the high redshift  type Ia
supernovae (Riess et al. 1998), and with previous limits on the
cosmological constant based on gravitational lensing (e.g., Falco et
al. 1998; Kochanek 1996; Chiba \& Yoshi 1997).

\section{SYSTEMATIC ERRORS}

\subsection{Errors Affecting the Expected Number of Lensed Galaxies}

{\it Errors in the photometric redshifts and in F}---Our
calculations rely on the accuracy of photometric redshifts, and hence
errors in these redshifts produce errors in the estimated number of lensed
galaxies.  The dispersion of redshifts in, e.g., the WBT
catalog with respect to spectroscopically measured ones, range from
0.03 to 0.1 for $z \lesssim 2$ and 0.14 to 0.36 for $z \gtrsim 2$. The
largest effect this could have on the expected number of lensed galaxies
would occur if the redshifts were {\em all} systematically low or high,
by an amount equal to the quoted error.
Even in such an extreme case, the 95\% confidence
lower limit on $\Omega_m-\Omega_{\Lambda}$ in our canonical case would
only range from -0.64 to -0.50, using respectively redshifts that are all
$1\sigma$ low and $1\sigma$ high compared to the best estimates in the
WBT catalog. This is actually a tremendous
overestimate of the effect of errors in the photometric redshifts. In
reality, a comparison of the photometric redshifts of WBT
with available spectroscopic redshifts indicates that the
errors are evenly distributed between high and low estimates, so the
overall expected number of lensed galaxies is barely affected.
The only systematic
effect visible in the WBT catalog is a paucity of
galaxies  in the redshift interval of 1.5 to 2.2. This has a small but
visible effect on our $\Omega_m-\Omega_\Lambda$ limits; the results for the WBT
catalog are larger than those of the SLY
catalog by $\sim$ 4\%. This increase is primarily due to the peaked
distribution of galaxies between redshifts of 2 and 2.5 in the WBT
catalog, while the same galaxies have a much broader
distribution in the SLY catalog. In addition, about 10
galaxies in the WBT catalog with $z \sim 2.5$ have
$z < 1$ in the SLY catalog.
Nonetheless, the two catalogs yield
expectations in the number of lensed galaxies that are
almost indistinguishable from each other.

Another possible systematic error has to do with the
way that the $F$ parameter (eq. [5]) is estimated. One method is
to use extensive local surveys of the galaxy luminosity function
and velocity dispersions to calculate $F$, then assume that because
most foreground lenses are at a redshift less than unity and
galaxy evolution out to that redshift is not believed to be
dramatic enough to change $F$ significantly (see Mao \& Kochanek 1994),
the value of $F$ to use in lensing calculations is the same as
it is locally.  This is the approach we adopt,
and it is supported by the luminosity  function in the HDF itself
(see \S\ 2.2).  Note, however, that
if $F$ were to be inferred solely from observations of a particular field,
such as the HDF, then because the inferred number density and luminosity
of galaxies depends on the assumed cosmology, so will $F$.  The
dependence of the inferred $F$ on cosmology also depends on the
relation between luminosity and velocity dispersion, but
if $\gamma=4$, the effect is that $F$ is smaller when $\Omega_\Lambda>0$
than when $\Omega_m=1$ and $\Omega_\Lambda=0$, as is often assumed.
If $\Omega_m-\Omega_\Lambda\sim -0.4$, the effect is to
decrease $F$ by $\sim 30$\% for lenses between $z=0.5$ and $z=1$.
If the field of interest contains a small enough number of potential
lenses so that fluctuations are important and $F$ must be derived from
that field, this effect may have to be included.  In the case of the
HDF, however, the number counts are large and the expected level
of fluctuations is small, so we assume that $F$ is the same as it is locally.

{\it Errors in calculating the number of lensed galaxies}---Another
source of systematic error in our study is that we have used the
analytical filled-beam expression to estimate the probability of lensing
at high redshift. This calculation underestimates the true probability based
on numerical techniques by about 2\% for the galaxy redshifts
of the HDF (Holz et al. 1999).  Thus, we have
underestimated the expected number of lensed galaxies in HDF.
This is a systematic error in our calculation,
and it implies that our lower limits
on $\Omega_m-\Omega_{\Lambda}$ have been underestimated by a similar amount.

We have also assumed that lensing galaxies can be described  by isothermal
spheres. However, it is likely that galaxies have a non-negligible
core  radius. Such a core radius can
decrease the expected  number of lensed galaxies present in the HDF
compared to the number calculated assuming a zero core radius.
Kochanek (1996) addressed
the issue of a finite core radius by studying lens models described by
softened isothermal spheres, and showed that a finite core radius
increases the velocity dispersion; thus, the decrease in lensing
probability due to an added core radius is compensated by the
increase in velocity dispersion.  In addition, the observed core
radii of E and S0 galaxies that dominate lensing are
much smaller than their Einstein radii, and hence the effect of a
finite core radius is small (Kochanek 1996).  We therefore do not
expect the effect of finite
galaxy core radii to dramatically change the lower limits on
$\Omega_m - \Omega_\Lambda$ presented in this paper.

\subsection{Errors Affecting the Observed Number of Lensed Galaxies}

{\it Reddening effects}---Given that the light from a lensed
galaxy must pass near or through a lensing galaxy, the column
depth of dust in the foreground of  multiply-imaged galaxies may be
systematically higher than for the field as a whole.  Hence,
the extinction of images could be of
considerable importance in lens studies.  Most lensing galaxies
are in a redshift range $z\sim 0.5-1.0$, so it is the properties
of galaxies at those redshifts that are most important in
evaluating the likely effect of extinction.

By comparing radio-selected and optical-selected lens samples,
Malhotra et al. (1997) suggested
that optical lens searches are heavily impaired by extinction due to
dust, with a mean magnitude change of 2 $\pm$ 1 magnitudes between
images.  A different conclusion was reached by Falco et al. (1998),
who suggested that extinction from dust only produces a mean magnitude
change of $\sim$ 0.5.  The conclusion of Falco et al.\ (1998) that
extinction has a minor impact on lensing studies is supported by
Kochanek (1996), who presented a reddening model in lensing
statistics and determined that extinction in lensing galaxies
decreases the probability of observing multiple lensing by only
$\sim$10\%.  This calculation concentrated on the $z \sim 0$
elliptical galaxies, which are thought to be fairly similar to the
galaxies at $z\sim 0.5-1.0$ that are doing most of the lensing,
although the inferred increase in the massive star formation rate
near $z\sim 1$ (e.g., Madau et al. 1998) may imply somewhat higher extinction
than at low redshifts.

Given that there is no detailed analysis of the dust distribution associated
with HDF galaxies, we have not included the effects of reddening due to dust
in the present calculation. However, our variation of the detection
threshold magnitude, $m_{\rm lim}$, in Table~2 is indicative of the
possible effects of extinction on our results.
We find that even a change as large as 1 magnitude
has a fairly minor effect on our limits.

{\it Source confusion and image separation effects}---The clustering
of optical galaxies may confuse lens search programs in two ways:
On the one hand, clustering can lead one to infer the existence of
multiply-lensed galaxies under the
assumption that observed galaxies are images of a lensed background object,
when in fact the images are unrelated; while, on the other
hand, clustering of galaxies
can confuse lensed-image identification
by increasing the surface brightness of the surrounding regions near the
images.
The two gravitationally lensed
sources found by Ratnatunga et al. (1995) using HST WFPC images
have image separations of the order $\sim$ 1.2$''$ to 1.5$''$.
For presently confirmed lenses, the effective diameter of the
foreground lensing galaxies at $z \sim 1$
are of the order $\sim$ 1.5$''$, similar to lensed image separations. Thus,
galaxy clustering may impair the detection of lensed sources.
This effect may in fact be present in the currently confirmed lensed
source sample, where almost all
of the lensed sources with small image separations were initially
selected in radio searches.
However, gravitationally lensed
galaxies typically have image shapes that can be differentiated
from random clustering of galaxies, and generally have colors different from
foreground galaxies. Therefore, a careful examination of the HDF,
including both colors and positions, should efficiently reveal lensed sources.
The recent discovery of about 10 new small separation lensed sources
in the HST Medium Deep Survey (K. Ratnatunga, private communication)
suggests that efficient lens searches can be and have been made.
Therefore, we do not
expect that a large number of HDF lenses have been missed,
due to galaxy clustering or source confusion.

\section{SUMMARY AND CONCLUSIONS}

By comparing the expected number of lensed galaxies in the HDF
to the observed number of lensed galaxies,
we have presented limits on the cosmological parameters. We have considered
the possibility that present lens search programs may have missed
one to three possible lensed galaxies in the HDF, and have given
our cosmological parameter constraints for a variety of
cases, including the change in limiting magnitude for lens searches.

We find that the expected number of lenses in the HDF is primarily
a function of $\Omega_m-\Omega_\Lambda$. A comparison of the
expected number of lensed galaxies with
the observed number allows us to put a lower limit on
the current value of $\Omega_m-\Omega_{\Lambda}$.
Making use of the
HDF luminosity function (as determined by SLY),
our 95\% confidence lower limit on
$\Omega_m-\Omega_{\Lambda}$ ranges between -0.44 and -0.73 (see Table~2).
If the only lensed galaxy in the HDF is the one candidate found
by Zepf et al. (1997), then $\Omega_m-\Omega_{\Lambda} > -0.58$.
These lower limits
are not in conflict with estimates based on high redshift supernovae
(viz., $\Omega_m - \Omega_\Lambda\sim -0.5\pm 0.4$ [Riess et al. 1998]).

As has been recently noted in the literature,
combining $\Omega_m-\Omega_{\Lambda}$ results from high redshift
supernovae measurements with
$\Omega_m+\Omega_{\Lambda}$ results from CMB power spectrum analysis
constrains $\Omega_m$ and $\Omega_{\Lambda}$ separately, with
much higher accuracy than the individual experiments alone (White 1998;
Eisenstein, Hu, \& Tegmark 1998). We note that gravitational
lensing constraints on $\Omega_m-\Omega_\Lambda$ should also be
considered in such an analysis.

We have shown, by comparing our results from two different catalogs,
that photometric redshifts can be used to estimate the expected number
of lensed galaxies in the HDF with reasonable accuracy.
This bodes well for the upcoming Southern Hubble Deep Field
redshift catalog that is expected in the near-future.
The Southern HDF will double the number of high redshift galaxies
and will thus double the expected number of gravitationally lensed galaxies.
If, for example, $\Omega_m-\Omega_{\Lambda}\sim -0.5$,
then from Table~1 there could be three lensed galaxies
detected in the Southern HDF.
The actual number of detected galaxies will lead to strong constraints
on $\Omega_m-\Omega_{\Lambda}$.

\acknowledgments
We thank Yun Wang, Neta Bahcall, Ed Turner and Marcin Sawicki
for making their photometric
redshift catalogs publicly available, and Stephen Zepf for
useful discussions regarding the observed number of lensed sources
in the HDF. We thank the anonymous referee for
constructive comments on the paper. This work was supported in part
by NASA grant NAG~5-2868 (MCM),
and by NASA grant NAG~5-4406 and NSF grant DMS~97-09696 (JMQ).
ARC acknowledges partial support from the
McCormick Fellowship at the University of Chicago,
and a Grant-In-Aid of Research from the National Academy of
Sciences, awarded through Sigma Xi, the Scientific Research Society.

\begin{figure*}[t]
\hbox{\hskip 0.0truein
\centerline{\psfig{file=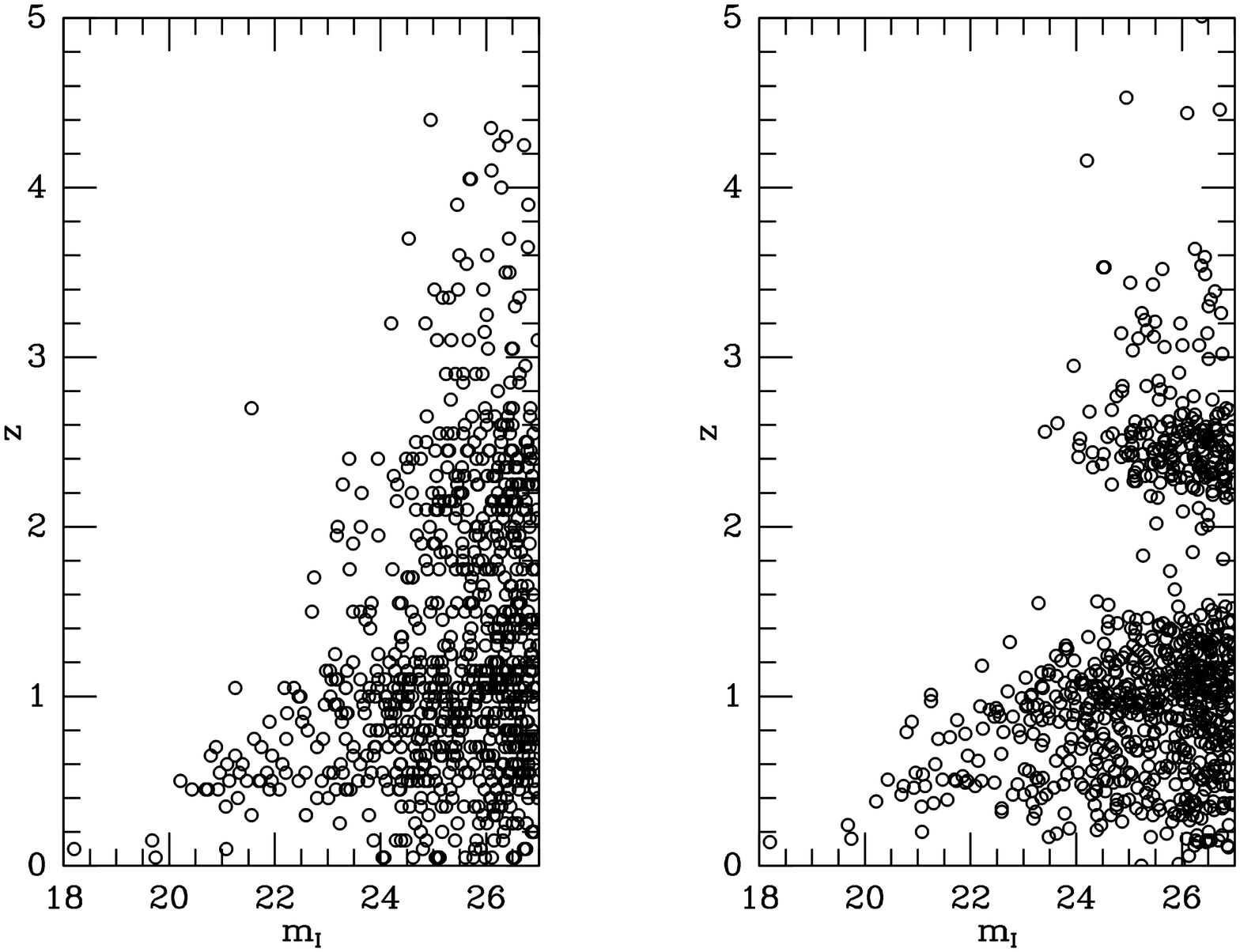,width=6.0truein,angle=-90}\hfil}}
\caption[fig1]{Redshift distribution of 848 galaxies with
I-band magnitude $\lesssim 27$ in the HDF. The plot shows the photometric
redshifts derived by Sawicki, Lee, \& Yin (1997; SLY) ({\em left panel}) and
Wang, Bahcall, \& Turner (1998; WBT) ({\em right panel}),
as a function of the I-band magnitude.
Both catalogs appear to trace the same redshift distribution, with two
peaks ($z \sim 0.6$ and 2.3). However, there is a lack of galaxies in the
WBT catalog between $z \sim 1.5$ to 2.2. This is the same range
in redshift where no spectroscopic redshifts are currently available for the
HDF.}
\end{figure*}

\begin{figure*}[t]
\hbox{\hskip 0.0truein
\centerline{\psfig{file=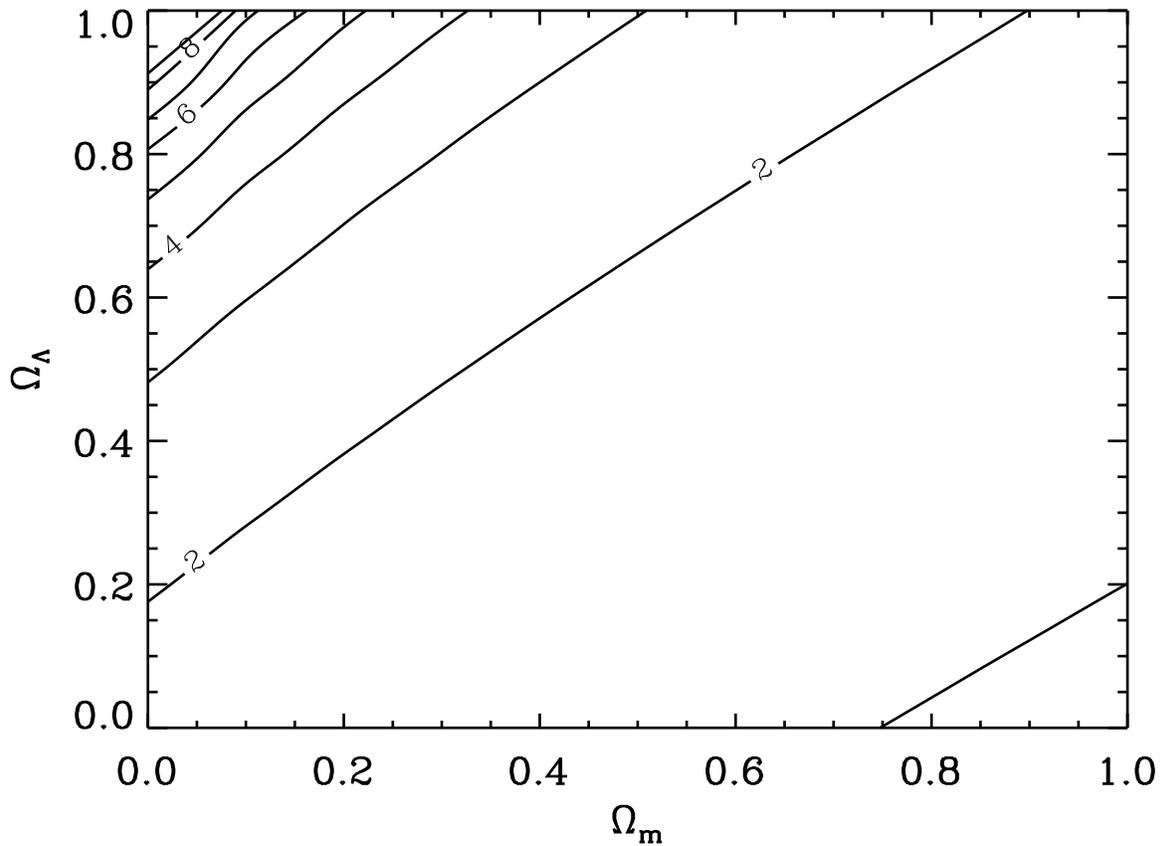,width=6.0truein,angle=90}\hfil}}
\caption[fig2]{Expected number of multiply-imaged galaxies, $\bar N$,
in the HDF, as a function of $\Omega_m$ and $\Omega_\Lambda$.
$\bar N$ is constant along lines of constant $\Omega_m-\Omega_\Lambda$,
allowing for direct constraints on this quantity.
Shown here is the expected number  based on the SLY catalog, and for
lens search programs that have been carried out to a limiting magnitude of
$m_{\rm lim}=28.5$ in HDF I-band images.}
\end{figure*}

\begin{figure*}[t]
\hbox{\hskip 0.0truein
\centerline{\psfig{file=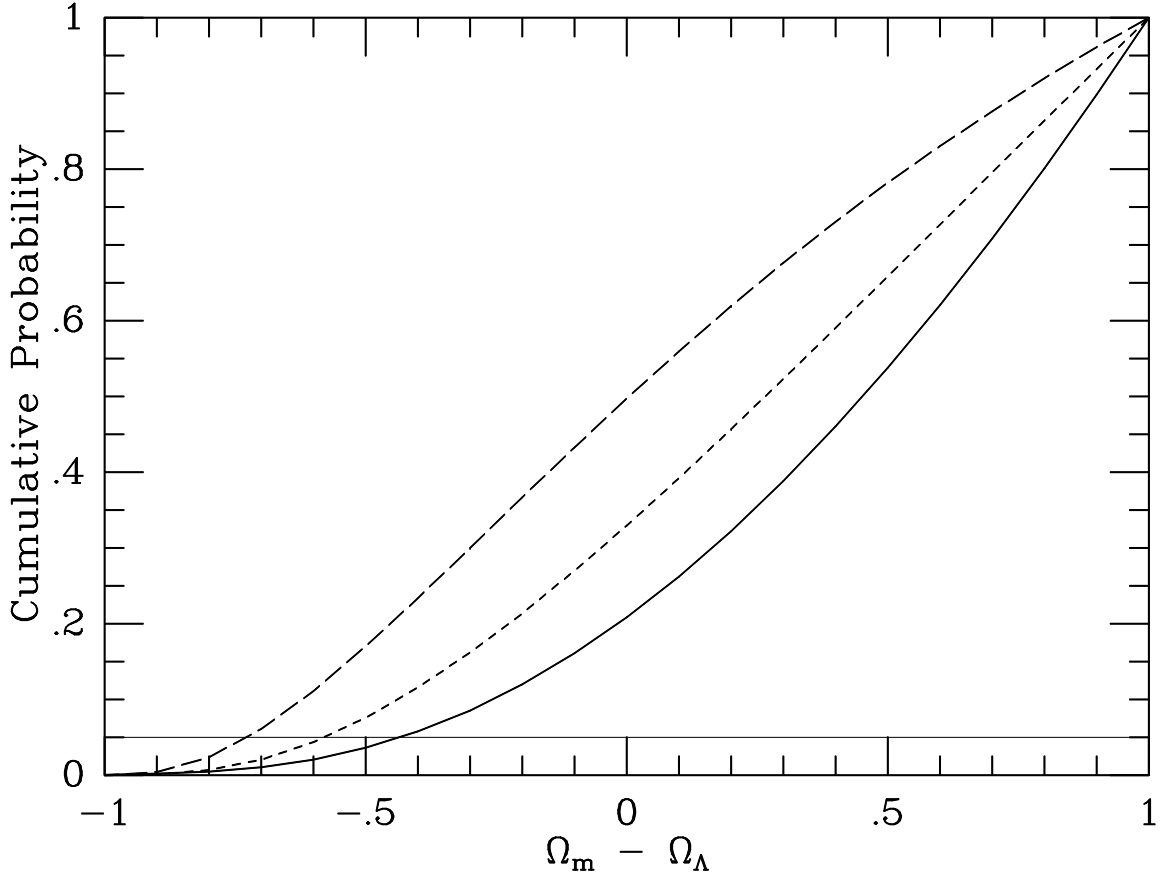,width=6.0truein,angle=-90}\hfil}}
\caption[fig3]{Cumulative probability distribution for
$\Omega_m-\Omega_{\Lambda}$,
if there is no lensed galaxy ({\em solid}),
one lensed galaxy ({\em short-dashed}),
and two lensed galaxies ({\em long-dashed}) in the HDF,
down to a lens-search limiting magnitude of $m_{\rm lim}=28.5$
in I-band images (using photometric redshifts from the SLY catalog).
The intercepts with the horizontal line show the 95\% confidence
lower limits on $\Omega_m-\Omega_{\Lambda}$ (see Table~2).}
\end{figure*}

\clearpage

\begin{deluxetable}{rcc}
\tablewidth{30pc}
\tablenum{1}
\tablecaption{Expected Number of Lensed Galaxies in the HDF.}
\tablehead{
\colhead{$\Omega_m-\Omega_\Lambda$} &
\colhead{Sawicki, Lee, \& Yin 1997}    &
\colhead{Wang, Bahcall, \& Turner 1998} }
\startdata
-1.0  & 14.3 &  15.8  \nl
-0.9 &  8.2 &  9.0 \nl
-0.8 &  5.4 &  5.8 \nl
-0.7 &  4.3 &  4.6 \nl
-0.6 &  3.5 &  3.7 \nl
-0.5 &  3.0 &  3.2 \nl
-0.4 &  2.6 &  2.8 \nl
-0.3 &  2.3 &  2.4 \nl
-0.2 &  2.0 &  2.2 \nl
-0.1 &  1.8 &  2.0 \nl
0.0  &  1.7 &  1.8 \nl
0.1  &  1.6 &  1.6 \nl
0.2  &  1.4 &  1.5 \nl
0.3  &  1.3 &  1.4 \nl
0.4 &   1.2 &  1.3 \nl
0.5 &   1.2 &  1.2 \nl
0.6 &   1.1 &  1.1 \nl
0.7 &   1.0 &  1.0 \nl
0.8 &   1.0 &  1.0 \nl
0.9 &   0.9 &  0.9 \nl
1.0 &   0.9 &  0.9 \nl
\enddata
\end{deluxetable}

\begin{deluxetable}{lcc}
\tablewidth{40pc}
\tablenum{2}
\tablecaption{95\% confidence lower limits on $\Omega_m-\Omega_{\Lambda}$.}
\tablehead{
\colhead{} &
\colhead{Sawicki, Lee, \& Yin 1997}    &
\colhead{Wang, Bahcall, \& Turner 1998} }
\startdata
$m_{\rm lim}=28.5$, $\alpha=-2.1$, $n=1$ $(z_s=2.5)$ & -0.58 &	-0.56 \nl
$m_{\rm lim}=28.5$, $\alpha=-2.1$, $n=0$ &	-0.44 &			-0.42
\nl
$m_{\rm lim}=28.5$, $\alpha=-2.1$, $n=1$ $(z_s=1.0)$ & -0.52 &	-0.49 \nl
$m_{\rm lim}=28.5$, $\alpha=-2.1$, $n=2$ $(z_s=2.5)$ &	-0.73 &		-0.70
\nl
$m_{\rm lim}=28.5$, $\alpha=-2.0$, $n=1$ $(z_s=2.5)$ & -0.64 & 	-0.62 \nl
$m_{\rm lim}=28.5$, $\alpha=-2.2$, $n=1$ $(z_s=2.5)$ & -0.56 &	-0.54 \nl
$m_{\rm lim} = 28.0$, $\alpha=-2.1$, $n=1$ $(z_s=2.5)$ & -0.63 &	-0.60
\nl
$m_{\rm lim} = 27.5$, $\alpha=-2.1$, $n=1$ $(z_s=2.5)$ & -0.70 &	-0.67
\nl
\enddata
\end{deluxetable}

\end{document}